\documentstyle[aps]{revtex}

\title
{A Path Integral Approach to Derivative Security Pricing:\\
I. Formalism and Analytical Results} 
\author{Marco Rosa-Clot and Stefano Taddei}
\address
{Dipartimento di Fisica, Universit\`a degli Studi di Firenze\\
and Istituto Nazionale di Fisica Nucleare, Sezione di
Firenze,\\
Largo Enrico Fermi 2, I-50125, Firenze, Italy}

\begin{document}

\maketitle

\begin{abstract} 

We use a path integral approach for solving the stochastic equations underlying
the financial markets, and
we show the equivalence between the path integral and the usual SDE and PDE
methods. 
We analyze both the one-dimensional and the multi-dimensional cases, with
point dependent drift and volatility, and describe
a covariant formulation which allows general changes of variables.
Finally we apply the method to some economic models with analytical solutions.
In particular, we evaluate the expectation value of functionals which
correspond to quantities of financial interest.

\end{abstract} 


\section*{Introduction}

The starting point of our analysis is that quantities as the stock prices,
the option prices, and the interest rates
satisfy differential stochastic equations (SDEs), i.e. ordinary 
differential equations with a superposed white noise.
Such equations are called Langevin equations, and they are extensively used
in the financial literature. The solutions of such equations are usually
obtained by solving the associated partial differential equations (PDEs).

In this paper we want to describe an alternative approach based on the
path integral formulation. The notion of path integrals, also called Wiener
integrals in stochastic calculus, and Feynman integrals in quantum mechanics,
is known from a long time. A proper mathematical definition 
of the Wiener integral can be found in the original works
of Wiener\cite{wiener}, and Kac\cite{kac}, while the quantum
mechanical analogous has been introduced by Feynman\cite{feyn}.

The importance of this formalism lies in the possibility of employing powerful
analytical and numerical techniques, developed in physics, for solving the
usual problems of option pricing. Some attempts of using this approach
in finance have been described in recent literature (see, for instance,
Ref.\cite{baaquie}). Here we discuss the path integral formulation in a
general manner, and, as examples, we solve some well known economical models.
Furthermore, in a forthcoming paper, we will describe some numerical methods.

In section \ref{s:1.10}, we show the general one-dimensional formalism.
In section \ref{s:multi},
we extend the formulation to the multi-dimensional case. In section
\ref{s:cov}, we
discuss a covariant formulation which is necessary to perform a general
transformation of variables. In section
\ref{s:exp}, we define the expectation value
of a general functional. Finally, in section
\ref{s:analytical}, we give some analytical results.

\section{Transition probability and path integral
formalism in one dimension}
\label{s:1.10}

\subsection{Langevin equation and discretization problem}  

In 1908 the french physicist Paul Langevin\cite{langevin} wrote down a
differential equation containing a Gaussian white noise 
coefficient, $f(\tau)$, 
\begin{equation}   
{\dot x}(\tau)=a(x,\tau)+\sigma(x,\tau)f(\tau). 
\label{e:1.1}  
\end{equation}   
where $x(\tau)$ is a stochastic process to be determined. Such 
white-noise-driven differential equations are often used
in physics and chemistry  and are the oldest form of SDEs.
The Langevin equation written above
does not define univocally a stochastic process, and it has to be 
supplemented with an additional interpretation rule
(see, for instance, Refs.\cite{arnold,lang,gard}). 
This is related to the ambiguity in the
discretizations of this equation.
We may remove this ambiguity by introducing 
additional structure into the Eq. (\ref{e:1.1}), and discretizing
the Langevin equation as 
\begin{equation} 
\Delta x=a(y+\zeta\Delta x,t)\Delta t+ 
\sigma(y+\eta\Delta x,t)\Delta w,
\label{e:1.2} 
\end{equation} 
where $t$ is the initial time, $\Delta t $ the time step, and $y=x(t)$.  
If we expand in $\Delta x$ and recall that
\hbox{$\Delta w\sim O(\sqrt{\Delta t})$}, we find,
\hbox{$\Delta x=\sigma(y,t)\; \Delta w+O(\Delta t)$}, and,
to the leading order in $\Delta t$,
\begin{equation} 
\Delta x=a(y,t)\;\Delta t+
\eta\;\sigma(y,t)\;{\partial\sigma\over\partial x}(y,t)\;\Delta w^2+
\sigma(y,t)\;\Delta w.
\label{e:1.3old} 
\end{equation} 
Now, since in the framework of stochastic calculus the
following equality holds,
\begin{equation} 
\Delta w^2{\dot =}\Delta t,
\label{e:1.dwdt}
\end{equation}
(where we used the usual symbol\cite{witt}, ${\dot =}$),
the Eq. (\ref{e:1.3old}) becomes
\begin{equation}
\Delta x=A(y,t)\;\Delta t+
\sigma(y,t)\;\Delta w,
\label{e:1.3}
\end{equation}
where
\begin{equation} 
A(x,\tau)=a(x,\tau)+
\eta\;\sigma(x,\tau)\;{\partial\sigma\over\partial x}(x,\tau).
\label{e:1.amu}
\end{equation}
The Langevin equation, written in this form, describes a well defined 
stochastic process,
and different $\eta$'s correspond to distinct processes.
For example, $\eta=0$ and $\displaystyle{\eta={1\over 2}}$ correspond to the
It\^o and the Stratonovich interpretations, respectively.
Note that the Eq. (\ref{e:1.3}) does not depend on $\zeta$.
In conclusion,
\begin{itemize}
\item[I.] {\em A stochastic differential equation is well defined only if both
a continuous expression and a discretization rule are given.}
\end{itemize}
From now on, we will always write the underlying stochastic equation in
the discretized form (\ref{e:1.3}), understanding that the continuous
limit must be taken. Therefore the stochastic process will be
defined when the functions $A(x,\tau)$ and $\sigma(x,\tau)$ are given.

Unfortunately there is still an ambiguity. In fact,
even if a stochastic process is well defined, the SDE
describing such a process is not univocal. Let us consider the following
family of equations, depending on the parameter, $\kappa$,
\begin{equation}   
{\dot x}(\tau)=
a(x,\tau)-\kappa\;\sigma(x,\tau)\;{\partial\sigma\over\partial x}(x,\tau)+
\sigma(x,\tau)f(\tau).
\label{e:1.1family}  
\end{equation}   
By expanding in $\Delta x$, as above, we obtain
\begin{equation} 
\Delta x=[a(y,t)+
(\eta-\kappa)\;\sigma(y,t)\;{\partial\sigma\over\partial x}(y,t)]\;\Delta t+
\sigma(y,t)\;\Delta w.
\label{e:1.3family} 
\end{equation} 
Therefore, if we choose the appropriate
discretization rule for each parameter, $\kappa$, i.e.
by fixing $\eta$ such that \hbox{$\eta-\kappa$} is fixed,
we can describe the same process by different SDEs. In other words,
\begin{itemize}
\item[II.] {\em Many different continuous expressions for the SDE, with the
appropriate discretization rule, define the same stochastic process.}
\end{itemize}

We want to stress that all economic and financial applications of
stochastic
calculus have used so far the It\^o prescription because it gives
the coefficients
of the SDE a simpler and meaningful interpretation (in particular, the
drift coefficient, $a(x,\tau)$, appears directly). On the other hand,
the Stratonovich prescription, for example,
allows to employ the usual rules of calculus
instead of the more complex It\^o calculus.
However we have seen above
that any SDE with the It\^o prescription is equivalent to some other SDE with
the appropriate prescription.
This more general approach is needed
in order to describe the connection with the equivalent difficulties which
arise in the path integral formulation, where we cannot avoid a general
treatment to explain the analytical and numerical techniques of
computation.

Finally, we observe that an alternative description of the stochastic
process is given by the following equivalent partial differential equations,
which can be derived
in an unambiguous way from the Langevin equation (\ref{e:1.3})
(see, for example, Ref.\cite{risken}):
\begin{itemize}
\item
Kolmogorov's backward equation 
\begin{equation}  
{\partial\over\partial t}\;\rho(x,T\mid y,t)=\left\{A(y,t)\;{\partial 
\over\partial y}+{1\over2}\;\sigma^2(y,t)\;{\partial^2\over\partial  
y^2}\right\}\;\rho(x,T\mid y,t);
\label{e:1.5}  
\end{equation} 
\item
Kolmogorov's forward equation or Fokker-Planck equation 
\begin{equation}  
{\partial\over\partial T}\;\rho(x,T\mid y,t)= \left\{-{\partial\over\partial 
x}\;A(x,T)+{1\over2}{\partial^2\over\partial x^2}\; 
\sigma^2(x,T)\right\}\; \rho(x,T\mid y,t).
\label{e:1.5a}
\end{equation} 
\end{itemize}
The solution of these equations is the transition probability function,
$\rho(x,T\mid y,t)$.
Note that the ambiguities of the continuous Langevin equation (\ref{e:1.1})
are not present here for the following reasons:
the first ambiguity is solved once the
functions $A(x,\tau)$ and $\sigma(x,\tau)$ are given;
the second one, which is connected to the differential
operator ordering, is fixed once this ordering has been fixed.

\subsection{Stochastic differential equation and
short-time transition probability}

The Brownian motion can be seen as the convolution of an infinite sequence 
of infinitesimal (short-time) steps. This constitutes the  bridge between
local equations and an integral formulation of the problem. 
Let us write the simple stochastic equation 
\begin{equation} 
\Delta x = \sigma \Delta w.
\label{eq:free} 
\end{equation}  
If $w$ is  a Wiener process this equation define a Markov process with zero 
average  and variance equal to $\sigma^2 \Delta t$. 
The corresponding short-time transition probability is given by  
\begin{equation} 
\rho(x,t+\Delta t \mid y,t)=\sqrt {1\over{2\pi\;\Delta t\;\sigma^2}}\; 
\exp \left\{ {{-(x-y)^2}\over{2 \;\Delta t\; \sigma^2}} \right\}.
\end{equation} 
In general, we do not have an explicit expression for
the short-time transition probability 
corresponding to the stochastic equation (\ref{e:1.3}).
However we can write the following general expression,
which is correct up to $O(\Delta t)$,
\begin{equation} 
\rho(x,t+\Delta t \mid y,t)\simeq \sqrt {1\over{2\pi\;\Delta t\;
{\sigma(y,t)}^2}}\; 
\exp \left\{ {{-(x-y-A(y,t) \;\Delta t)^2}\over{2 \;\Delta t \;
{\sigma(y,t)}^2}} \right\},
\label{e:onepre}
\end{equation}
where $A(x,\tau)$ is given in Eq. (\ref{e:1.amu}).
A proof of this is reported for completeness in Appendix \ref{a:a}.
The equation (\ref{e:onepre}) gives a prescription to write
the solution of a Fokker-Planck 
equation in the form of a convolution product of short-time transition
probability functions.

\subsection{Finite time transition probability and path integral}

The finite time transition probability can be written
as the convolution of short-time transition probabilities
\begin{eqnarray}
\rho(x,T \mid y,t)\;=\int_{-\infty}^{+\infty}\dots
\int_{-\infty}^{+\infty}&&dx_Ndx_{N-1}\dots dx_1\;\;
\rho(x,T \mid x_N,T-\Delta t)\nonumber\\
&&\;\;\;\;\rho(x_N,T-\Delta
t \mid x_{N-1},T-2\Delta t)\dots\rho(x_1,t+\Delta t \mid y,t),
\label{e:discret}
\end{eqnarray}
with $\Delta t = \displaystyle{T-t\over N}$.
Therefore, by substituting the expression (\ref{e:onepre}) in the previous
equation, we get
\begin{equation}
\rho(x,T\mid y,t)\;\simeq \int_{-\infty}^{+\infty}\!\!\!\!\!\!\!\!\!\ldots\!
\int_{-\infty}^{+\infty}
\prod^{N}_{i=1}dx_i
\prod^{N+1}_{j=1}
{\sigma(x_{j-1},t_{j-1})^{-1}\over\sqrt{2\pi\,\Delta t}}\;
\exp\,\Bigl\{-S_0(x_j,t_j;x_{j-1},t_{j-1})\Bigr\},
\label{e:dis1}
\end{equation}
with
\begin{equation} 
S_0(x_j,t_j;x_{j-1},t_{j-1})={1\over 
2\;\sigma(x_{j-1},t_{j-1})^2}\;
\left[{(x_j-x_{j-1})\over\Delta t}- 
A(x_{j-1},t_{j-1})\right]^2\;\Delta t
\label{e:dis2} 
\end{equation} 
(the meaning of the subscript, $0$, will be explained in the next section).
We can interpret $\displaystyle{{(x_j-x_{j-1})}\over{\Delta t}}$
as a mean velocity in the time 
interval $\Delta t$, and  $\sigma(x_{j-1},t_{j-1})^{-2}$ as a mass. 
Then a Lagrangian structure appears explicitly.
Hence, the limit for $N\rightarrow\infty$ of the RHS of Eq. (\ref{e:dis1})
can be formally written as
\begin{equation}
\rho(x,T\mid y,t)={\int\!\!\!\int}_{x(t)=y}^{x(T)=x}
{\cal D}[\sigma(x,\tau)^{-1}x(\tau)]\;
\exp\left\{-\int_t^T L_0[x(\tau),{\dot
x}(\tau);\tau]\;d\tau\right\}.
\label{e:path}
\end{equation}
The functional measure, ${\cal D}[\sigma(x,\tau)^{-1}x(\tau)]$, means
summation on
all possible paths starting from $x(t) = y$ and arriving at $x(T)=x$.
The integral to the exponent must be interpreted as the limit of the
discrete summation for $N\rightarrow\infty$, and the function
$L_0[x(\tau),{\dot x}(\tau);\tau]$ is a Lagrangian function defined by
\begin{equation}
L_0[x,{\dot x};\tau]=
{1\over 2\;\sigma(x,\tau)^2}\;
[{\dot x}-A(x,\tau)]^2.
\label{eq:lagr}
\end{equation}
The RHS of Eq. (\ref{e:path}) is called ``path integral''.
Since in the limit $N\rightarrow\infty$ only $O(\Delta t)$ contribute to
the integral, the formal expression in Eq. (\ref{e:path}) gives the exact
finite time transition probability.
In conclusion, there is a complete equivalence between the differential
stochastic equation,
the Fokker-Planck equation and the path integral approach.

Note that in the limit $N\rightarrow\infty$ distinct expressions
for the short-time transition probability, equal up to $O(\Delta t)$,
give rise to the same functional expression for
the path integral; therefore orders greater than $O(\Delta t)$ do not
have any effects from an analytical point of view. However, if the path
integral is solved numerically,
a more accurate approximation for the
short-time transition probability can be useful.
A general procedure to obtain an approximation to any order is
described in Appendix \ref{a:b}.

\subsection {Path integral and discretization problem}
\label{s:discretization}

We note that the path integrals present exactly the same ambiguities of the
stochastic differential equations. In particular, it is clear from its
definition that the formal expression (\ref{e:path}) is well defined
only if a discretization rule, i.e. a short-time transition probability,
is also given, since in general different discretization rules give rise
to different
results. Therefore we can say,
\begin{itemize}
\item[III.] {\em A path integral is well defined only if both
a continuous formal expression and a discretization rule are given
{\rm (cf. with I)}.}
\end{itemize}

A further ambiguity is due to the fact that
different Lagrangians, discretized by the appropriate
rule, give rise to equivalent expressions for
the short-time transition probability.
For example let us consider the Lagrangian
\begin{equation} 
L_{1/2}[x,{\dot x};\tau]=
L_0[x,{\dot x};\tau]+{1\over 2}\;\partial_x A(x,\tau),
\label{e:lbar} 
\end{equation} 
with the discretization rule given by
\begin{equation} 
\rho(x,T\mid y,t)\!\simeq\!\int_{-\infty}^{+\infty}\!\!\!\!\!\!\!\!\!\ldots\! 
\int_{-\infty}^{+\infty}\! \prod^{N}_{i=1}dx_i \prod^{N+1}_{j=1} 
{\sigma(x_{j-1},t_{j-1})^{-1}\over\sqrt{2\pi\,\Delta t}}\; 
\exp\left\{-S_{1/2}(x_j,t_j;x_{j-1},t_{j-1})\right\},
\label{e:1.8bis} 
\end{equation} 
where
\begin{eqnarray} 
&&S_{1/2}(x_j,t_j;x_{j-1},t_{j-1})={\Delta t\over 
2\;\sigma(x_{j-1},t_{j-1})^2}\;\nonumber\\ 
&&\;\;\;\;\;\;\;\;\times\left[{(x_j-x_{j-1})\over\Delta t}- 
A({x_j+x_{j-1}\over 2},{t_j+t_{j-1}\over 2})
\right]^2+{\Delta t\over 2}\;
\partial_x A({x_j+x_{j-1}\over 2},{t_j+t_{j-1}\over 2}). 
\label{e:1.9bis} 
\end{eqnarray} 
By expanding in series of $\Delta x$ and $\Delta t$ up to $O(\Delta t)$,
we obtain
\begin{eqnarray} 
S_{1/2}(x_j,t_j;x_{j-1},t_{j-1})=&&{\Delta t\over 
2\;\sigma(x_{j-1},t_{j-1})^2}\;
\left[{(x_j-x_{j-1})\over\Delta\tau}-
A(x_{j-1},t_{j-1})\right]^2\nonumber\\
&&\;\;\;+{\Delta t\over 2}\;\partial_x A(x_{j-1},t_{j-1})-
{\Delta x^2\over 2\;\sigma(x_{j-1},t_{j-1})^2}\;\partial_x
A(x_{j-1},t_{j-1}). 
\label{e:1.9tris} 
\end{eqnarray} 
Then by using the following identity, analogous to the (\ref{e:1.dwdt}),
\begin{equation} 
\Delta x^2\;{\dot =}\;\sigma(x_{j-1},t_{j-1})^2\;\Delta t,
\label{e:1.dxdt}
\end{equation}
the Eqs. (\ref{e:1.8bis}) and (\ref{e:1.9bis}) become equal to the Eqs.
(\ref{e:dis1}) and (\ref{e:dis2}).
The formal expression of the path integral is
\begin{equation}
\rho(x,T\mid y,t)={\int\!\!\!\int}_{x(t)=y}^{x(T)=x}
{\cal D}[\sigma(x,\tau)^{-1}x(\tau)]\;
\exp\left\{-\int_t^T L_{1/2}[x(\tau),{\dot
x}(\tau);\tau]\;d\tau\right\}.
\label{e:pathmp}
\end{equation}
We will call {\em pre-point} the path integral
formulation given by the Eqs. (\ref{e:dis1}), (\ref{e:dis2}),
(\ref{e:path}), and (\ref{eq:lagr}),
and {\em mid-point} that given by the
Eqs. (\ref{e:lbar}), (\ref{e:1.8bis}), (\ref{e:1.9bis}),
and (\ref{e:pathmp}).
In conclusion,
\begin{itemize}
\item[IV.] {\em Many different continuous formal expressions
for the path integral, with the
appropriate discretization rule, define the same stochastic process
{\rm (cf. with II)}.}
\end{itemize}

The pre-point formulation is usually simpler to perform numerical computations,
while the mid-point one has some advantages to carry on analytical
calculations,
as we will see in the following. From now on, if not explicitly specified, we
will use the mid-point formulation, but for the sake of simplicity we will
omit the subscript $1/2$.
 
\subsection {Gauge transformation} 
\label{s:gauge} 

Let us consider the case  
\begin{equation}  
L[x,{\dot x};\tau]= {1\over 2\;\sigma^2}\;[{\dot x}-a(x)]^2+
{1\over 2}\;\partial_x a(x),
\end{equation} 
where we have taken
\hbox{$\sigma(x,\tau)=\sigma={\rm constant}$}, and $A(x,t)=a(x)$. 
The finite time transition probability is given by
\begin{eqnarray} 
\lefteqn{\rho(x,T\mid y,t)=}\nonumber\\ 
&&\;\;\;\;\;={\int\!\!\!\int}_{x(t)=y}^{x(T)=x}\! 
{\cal D}[\sigma^{-1}x(\tau)] 
\exp\!\left\{\!-\!\int_t^T\! 
\left[{1\over 2\sigma^2}\,{\dot x}^2- 
{1\over \sigma^2}\,a(x)\,{\dot x}+ 
{1\over 2\sigma^2}\,a(x)^2+ 
{1\over 2}\;\partial_x a(x)\right] 
d\tau\right\}.
\label{e:1.ex8b} 
\end{eqnarray} 
This expression
contains a coupling between the derivative of the stochastic variable,
${\dot x}$, and the term $a(x)$.
Let us perform a 
{\em gauge transformation} by introducing a function $\phi (x)$ such that 
\begin{equation} 
a(x)={d\phi\over dx};
\end{equation} 
then we have 
\begin{equation} 
{1\over\sigma^2}\int_t^T a(x)\,{\dot x}\;d\tau = 
{1\over\sigma^2}\int_t^T {d\phi\over dx}\; dx = 
{1\over\sigma^2}\;[\phi(x)-\phi(y)], 
\label{e:1.ex11}
\end{equation} 
which is the same for all paths and depends only on the end 
points.

It is important to note that the integrals in (\ref{e:1.ex11}) are
stochastic integrals (${\dot x}$ is not defined for 
a Brownian path), and the result (\ref{e:1.ex11}) is true only if we use 
the mid-point formulation (Stratonovich prescription).
In fact, if we use the pre-point formulation, the term 
\hbox{$\displaystyle{{1\over 2}\;\partial_x a(x)}$} in Eq. (\ref{e:1.ex8b})
should be dropped, while
\begin{eqnarray} 
{1\over\sigma^2}\int_t^T a(x)\,{\dot x}\;d\tau &=& 
{1\over\sigma^2}\int_t^T {d\phi\over dx}\; dx\nonumber\\ 
&=&{1\over\sigma^2}\;[\phi(x)-\phi(y)]-{1\over 2}\;\int_t^T 
{d\over dx^2} \phi(x(\tau))\;d\tau 
\end{eqnarray} 
(It\^o prescription). Of course the final result is the same in both 
cases.

The term in Eq. (\ref{e:1.ex11}) can be put out of the path integral, and 
Eq. (\ref{e:1.ex8b}) becomes 
\begin{eqnarray} 
\lefteqn{\rho(x,T\mid y,t)}\nonumber\\ 
&&=e^{\Delta\phi/\sigma^2} 
{\int\!\!\!\int}_{x(t)=y}^{x(T)=x} 
{\cal D}[\sigma^{-1}x(\tau)]\; 
\exp\left\{-\int_t^T 
\left[{1\over 2\sigma^2}\,{\dot x}^2+ 
{1\over 2\sigma^2}\,a(x)^2+ 
{1\over 2}\;\partial_x a(x)\right] 
d\tau\right\},\nonumber\\ 
\end{eqnarray} 
where 
\begin{equation} 
\Delta\phi=\int_t^T a(x)\;dx.
\end{equation}

In conclusion, the previous discussion shows that the convenience of adopting
the mid-point formulation is related to the possibility of using the
usual rules of integral calculus.

\section{Multi-dimensional case}
\label{s:multi}

Until now we have discussed only the one-dimensional case. 
In the general case of a multi-dimensional Langevin equation in
$n$ dimensions, we have a 
straightforward generalization of it. 
Let us write the discretized Langevin equation in the following form, 
\begin{equation} 
\Delta x^\mu=A^\mu({\bf x}(t),t)\;\Delta t+ 
\sigma^\mu_i({\bf x}(t),t)\;\Delta w_i 
\label{e:1.3corr} 
\end{equation} 
(the sum over repeated indices has been understood), where 
\begin{equation} 
A^\mu({\bf x},\tau)=a^\mu({\bf x},\tau)+\eta\;\sigma^\nu_i({\bf 
x},\tau)\; {\partial\over\partial x^\nu}\sigma^\mu_i({\bf x},\tau). 
\label{e:1.4corr} 
\end{equation}
The Eq. (\ref{e:1.3corr}) corresponds to the following 
Fokker-Planck equation (in the sense that they describe the same 
stochastic process) 
\begin{equation} 
{\partial\over\partial T}\;\rho({\bf x},T\mid {\bf y},t)= 
\left\{-{\partial\over\partial x^\mu}\;A^\mu({\bf x},T)+ 
{1\over 2}{\partial^2\over\partial 
x^\mu\partial x^\nu}\;G^{\mu\nu}({\bf x},T)\right\}\; \rho({\bf 
x},T\mid{\bf y},t), 
\label{e:1.5corr} 
\end{equation} 
where the differential operators in the right hand side act also on 
$\rho$, and 
\begin{equation} 
G^{\mu\nu}({\bf x},\tau)=\sigma^\mu_i({\bf x},\tau)\; 
\sigma^\nu_i({\bf x},\tau). 
\label{e:s3corr} 
\end{equation} 
The short-time transition probability is given by 
\begin{equation} 
\rho({\bf x}_j,t_j\mid {\bf x}_{j-1},t_{j-1})\simeq
(2\pi\Delta t)^{-n/2}\;\sqrt{G({\bf x}_{j-1},t_{j-1})}\; 
\exp\left\{-S_0({\bf x}_j,t_j;{\bf x}_{j-1},t_{j-1})\right\} 
\label{e:1.8corr} 
\end{equation} 
where 
\begin{eqnarray} 
\lefteqn{S_0({\bf x}_j,t_j;{\bf x}_{j-1},t_{j-1})}\nonumber\\ 
&&\;\;\;\;\;\;\;\;\;\;={1\over 2\,\Delta t}\; 
G_{\mu\nu}({\bf x}_{j-1},t_{j-1}) 
\left[\Delta x^\mu\!-\!A^\mu({\bf x}_{j-1},t_{j-1})\Delta t\right] 
\left[\Delta x^\nu\!-\!A^\nu({\bf x}_{j-1},t_{j-1})\Delta t\right],\nonumber\\ 
\label{e:1.9corr} 
\end{eqnarray} 
and ${\bf x}_0={\bf y}$, ${\bf x}_{N+1}={\bf x}$;
while a path integral representation of the finite time transition
probability is
\begin{equation} 
\rho({\bf x},T\mid {\bf y},t)={\int\!\!\!\int}_{{\bf x}(t)={\bf 
y}}^{{\bf x}(T)={\bf x}} {\cal D}[\sqrt{G}\;{\bf 
x}(\tau)]\; \exp\left\{-\int_t^T L_0[{\bf x}(\tau),{\dot 
{\bf x}}(\tau);\tau]\;d\tau\right\},
\label{e:multi} 
\end{equation} 
with
\begin{equation} 
L_0[{\bf x},{\dot {\bf x}};\tau]= 
{1\over 2}\;G_{\mu\nu}({\bf x},\tau)\; 
[{\dot x}^\mu-A^\mu({\bf x},\tau)]\, 
[{\dot x}^\nu-A^\nu({\bf x},\tau)],
\label{e:multil} 
\end{equation} 
and the discretization rule given above.
This expression for the transition probability corresponds to  
a well defined discretization procedure, i.e. the pre-point rule
(cf. with Eqs. (\ref{e:dis1}), (\ref{e:dis2}),
(\ref{e:path}), and (\ref{eq:lagr})). Another representation is given by the
continuous formal expression
\begin{equation} 
\rho({\bf x},T\mid {\bf y},t)={\int\!\!\!\int}_{{\bf x}(t)={\bf 
y}}^{{\bf x}(T)={\bf x}} {\cal D}[\sqrt{G}\;{\bf 
x}(\tau)]\; \exp\left\{-\int_t^T L_{1/2}[{\bf x}(\tau),{\dot 
{\bf x}}(\tau);\tau]\;d\tau\right\},
\label{e:multimp} 
\end{equation} 
with the Lagrangian
\begin{equation} 
L_{1/2}[{\bf x},{\dot {\bf x}};\tau]= 
L_0[{\bf x},{\dot {\bf x}};\tau]+
{1\over 2}\;{\partial\over\partial x^\mu}\;A^\mu({\bf x},\tau),
\label{e:multilmid} 
\end{equation} 
and the (mid-point) discretization rule
\begin{equation} 
\rho({\bf x}_j,t_j\mid {\bf x}_{j-1},t_{j-1})\simeq
(2\pi\Delta t)^{-n/2}\;\sqrt{G({\bf x}_{j-1},t_{j-1})}\; 
\exp\left\{-S_{1/2}({\bf x}_j,t_j;{\bf x}_{j-1},t_{j-1})\right\} ,
\label{e:1.8corrmid} 
\end{equation} 
where 
\begin{eqnarray} 
\lefteqn{S_{1/2}({\bf x}_j,t_j;{\bf x}_{j-1},t_{j-1})=
{\Delta t\over 2}\; 
G_{\mu\nu}({\bf x}_{j-1},t_{j-1})}\nonumber\\
&&\;\;\;\;\;\;\;\;\;\;\times
\left[{\Delta x^\mu\over \Delta t}\!-\!A^\mu({{\bf x}_j+{\bf x}_{j-1}\over 2},
{t_j+t_{j-1}\over 2})\right] 
\left[{\Delta x^\nu\over \Delta t}\!-\!A^\nu({{\bf x}_j+{\bf x}_{j-1}\over 2},
{t_j+t_{j-1}\over 2})\right]\nonumber\\
&&\;\;\;\;\;\;\;\;\;\;+{\Delta t\over 2}\;{\partial\over\partial x^\mu}\;
A^\mu({{\bf x}_j+{\bf x}_{j-1}\over 2},{t_j+t_{j-1}\over 2})
\label{e:1.9corrmid} 
\end{eqnarray} 
(cf. with Eqs. (\ref{e:1.8bis}) and (\ref{e:1.9bis})).

\section{Covariant formulation}
\label{s:cov}

The path integral formulations given in the previous sections are not
covariant. As a result, a general (non linear) 
transformation of variables in the path integral cannot be performed according 
to the usual rules of calculus.
In the following we describe a covariant formulation of path 
integrals\cite{graham,lrtbook}, and we use this formulation to solve 
some specific problems. 

The tensor, $G^{\mu\nu}$, defined in
Eq. (\ref{e:s3corr}), transforms under a general
transformation of variables as a contravariant tensor
(see, for instance, Ref.\cite{lrtbook}). If it is invertible 
-- we will assume that this is always the case -- it can be interpreted as 
the metric tensor of a Riemannian manifold. On the other hand,
$A^\mu$ is not a contravariant vector, while the quantity 
\begin{equation} 
h^\mu({\bf x},\tau)=A^\mu({\bf x},\tau)- {1\over 
2\sqrt{G}}\;{\partial\over\partial x^\mu}\; \sqrt{G}\;G^{\mu\nu}({\bf 
x},\tau), 
\end{equation} 
with
\begin{equation} 
G\equiv G({\bf x},\tau)=\det G_{\mu\nu}({\bf x},\tau), 
\end{equation} 
transforms exactly as a contravariant vector.
Therefore a covariant path integral representation of the solution of
Eq. (\ref{e:1.5corr}) is given by 
\begin{equation} 
\rho({\bf x},T\mid {\bf y},t)={\int\!\!\!\int}_{{\bf x}(t)={\bf 
y}}^{{\bf x}(T)={\bf x}} {\cal D}[\sqrt{G}\;{\bf 
x}(\tau)]\; \exp\left\{-\int_t^T {\cal L}[{\bf x}(\tau),{\dot 
{\bf x}}(\tau);\tau]\;d\tau\right\}, 
\label{e:cov2} 
\end{equation} 
where 
\begin{equation} 
{\cal L}[{\bf x},{\dot {\bf x}};\tau]= 
{1\over 2}\;G_{\mu\nu}\; 
[{\dot x}^\mu-h^\mu]\, 
[{\dot x}^\nu-h^\nu]+ 
{1\over 2\sqrt{G}}\;{\partial\over\partial x^\mu}\; 
\sqrt{G}\;h^\mu+{1\over 12}\;R. 
\label{e:cov3} 
\end{equation} 
The scalar $R$ is the curvature 
\begin{equation} 
R=G^{\mu\nu}\;R^\lambda_{\phantom{\lambda}\mu\lambda\nu}\;, 
\label{e:r1} 
\end{equation} 
and 
\begin{equation} 
R^\lambda_{\phantom{\lambda}\mu\delta\nu}= 
{\partial\Gamma^\lambda_{\mu\delta}\over\partial x^\nu}- 
{\partial\Gamma^\lambda_{\mu\nu}\over\partial x^\delta}+ 
\Gamma^\eta_{\mu\delta}\Gamma^\lambda_{\nu\eta}- 
\Gamma^\eta_{\mu\nu}\Gamma^\lambda_{\delta\eta}\;, 
\label{e:r2} 
\end{equation} 
\begin{equation} 
\Gamma^\lambda_{\mu\nu}={1\over 2}\;G^{\delta\lambda}\; 
\left({\partial G_{\nu\delta}\over\partial x^\mu}+ 
{\partial G_{\mu\delta}\over\partial x^\nu}- 
{\partial G_{\nu\mu}\over\partial x^\delta}\right).
\label{e:r3} 
\end{equation} 
As usual, the continuous formal expression (\ref{e:cov2}) must be 
interpreted as the limit of a discretized expression 
\begin{equation} 
\rho({\bf x},T\mid {\bf y},t)=\int_{-\infty}^{+\infty} 
\!\!\!\!\!\!\!\!\!\ldots\! 
\int_{-\infty}^{+\infty} \prod^{N}_{i=1}d^nx_i \prod^{N+1}_{j=1} 
\rho({\bf x}_j,t_j\mid {\bf x}_{j-1},t_{j-1}).
\label{e:cov4} 
\end{equation} 
In general, however, even if a discretization rule is correct
in a given coordinate system, i.e.
the corresponding short-time transition probability satisfies the
Eq.\ (\ref{e:1.5corr}) up to $O(\Delta t)$,
after a transformation of variables it is not correct any more;
but a covariant path integral formulation needs also a covariant
discretization rule. A covariant discretization of the continuous
expression (\ref{e:cov2}) is given by\cite{graham,lrtbook}
\begin{eqnarray} 
\lefteqn{\rho({\bf x}_j,t_j\mid {\bf x}_{j-1},t_{j-1})\simeq
(2\pi\Delta t)^{-n/2}\;\sqrt{G({\bf x}_j,{\bar t}_j)}\; 
\exp\left\{-{1\over 2\Delta t}G_{\mu\nu}({\bf 
x}_{j-1},{\bar t}_j) \Delta x^\mu\Delta x^\nu\right\}}\nonumber\\ 
&&\;\;\;\;\;\;\times\;[1+G_{\mu\nu}h^\mu\Delta x^\nu+\Delta t (-{1\over 
2}G_{\mu\nu}h^\mu h^\nu- {1\over 2\sqrt{G}}\;{\partial\over\partial 
x^\mu}\;\sqrt{G}\;h^\mu -{1\over 12}R)\nonumber\\ 
&&\;\;\;\;\;\;\;\;\;\;\;\;\;+({1\over 2}\;(\partial_\mu G_{\lambda\nu}h^\lambda)-{1\over 
12}R_{\mu\nu})\Delta x^\mu\Delta x^\nu\nonumber\\ 
&&\;\;\;\;\;\;\;\;\;\;\;\;\;-{1\over 4}\;(\partial_\mu G_{\nu\lambda})\;\Delta x^\mu\Delta 
x^\nu\Delta x^\lambda\nonumber\\ 
&&\;\;\;\;\;\;\;\;\;\;\;\;\;-({1\over 12}\;(\partial_\mu\partial_\nu G_{\alpha\beta})- 
{1\over 24}\Gamma_{\mu\lambda\nu} 
\Gamma_{\alpha\phantom{\lambda}\beta}^{\phantom{\alpha}\lambda}) 
\Delta x^\mu\Delta x^\nu\Delta x^\alpha\Delta x^\beta/\Delta t
\nonumber\\ 
&&\;\;\;\;\;\;\;\;\;\;\;\;\;+{1\over 2}(-G_{\mu\nu}h^\mu\Delta x^\nu+{1\over 4}\;(\partial_\mu 
G_{\nu\lambda})\;\Delta x^\mu\Delta x^\nu\Delta 
x^\lambda/\Delta t)^2], 
\label{e:cov5} 
\end{eqnarray} 
where ${\bar t}_j=(t_j+t_{j-1})/2$ and all functions, if it is 
not explicitly specified, are evaluated at ${\bf x}_{j-1}$ and at any 
time between $t_{j-1}$ and $t_j$. \\
With this definition of path integral, the formal measure in the 
expression (\ref{e:cov2}) corresponds to the limit 
\begin{equation} 
{\cal D}[\sqrt{G}\;{\bf x}(\tau)]=\lim_{\Delta t\rightarrow 0}\; 
\sqrt{{G({\bf x},{\bar t}_{N+1})\over(2\pi\Delta t)^n}}\; 
\prod_{i=1}^{N}\sqrt{{G({\bf x}_i,{\bar t}_i)\over 
(2\pi\Delta t)^n}}\;d^nx_i. 
\label{e:mes1} 
\end{equation} 
Therefore, since the quantity $\sqrt{G}\;d^nx$ is an invariant, after 
a change of variables the measure (\ref{e:mes1}) transforms in the 
following way 
\begin{equation} 
{\cal D}[\sqrt{G^\prime}\;{\bf x^\prime}(\tau)]=\left| 
{\partial x^\prime_\mu\over\partial x_\nu}\right|^{-1}_{{\bf 
x},T}{\cal D}[\sqrt{G}\;{\bf x}(\tau)], 
\label{e:mes2} 
\end{equation} 
where $\displaystyle{\left|{\partial x^\prime_\mu\over\partial 
x_\nu}\right|}_{{\bf x},T}$ is the Jacobian of the 
transformation calculated in the final point
(note that the symbol, ${}^\prime$, does not mean derivation).

\subsection{The case of $G^{\mu\nu}={\rm constant}$}
\label{s:gconst}

If $G^{\mu\nu}$ is constant, the
affine connections, $\Gamma^\lambda_{\mu\nu}$, and the curvature,
$R$, are equal to zero. Then the Lagrangian (\ref{e:cov3}) becomes
\begin{equation} 
{\cal L}[{\bf x},{\dot {\bf x}};\tau]= 
{1\over 2}\;G_{\mu\nu}\; 
[{\dot x}^\mu-A^\mu({\bf x},\tau)]\, 
[{\dot x}^\nu-A^\nu({\bf x},\tau)]+ 
{1\over 2}\;{\partial\over\partial x^\mu}\;A^\mu({\bf x},\tau)
\label{e:cov3const} 
\end{equation}
(cf. with Eq. (\ref{e:multilmid})),
and the discretized expression (\ref{e:cov5}),
\begin{eqnarray}
\lefteqn{\rho({\bf x}_j,t_j\mid {\bf x}_{j-1},t_{j-1})\simeq
(2\pi\Delta t)^{-n/2}\;\sqrt{G}\;
\exp\left\{-{1\over 2\Delta t}G_{\mu\nu}\Delta x^\mu\Delta
x^\nu\right\}}\nonumber\\
&&\;\;\;\;\times\;[1+G_{\mu\nu}A^\mu\Delta
x^\nu+\Delta t (-{1\over 2}G_{\mu\nu}A^\mu A^\nu- {1\over
2}\;{\partial\over\partial x^\mu}\;A^\mu)\nonumber\\
&&\;\;\;\;\;\;\;\;\;\;\;+{1\over 2}\;G_{\lambda\nu}\Delta x^\mu\Delta x^\nu
{\partial\over\partial x^\mu}\;A^\lambda+
{1\over 2}\;(-G_{\mu\nu}A^\mu\Delta x^\nu)^2].
\label{e:cov5bis}
\end{eqnarray}
Note that the equation (\ref{e:cov5bis}) is simply the expansion up to
$O(\Delta t)$ of the expression in (\ref{e:1.8corrmid}).

In conclusion, in the case of $G^{\mu\nu}={\rm constant}$, the covariant
formulation becomes the mid-point one.
Therefore, if we can make a change of variables which gives a constant metric
(f\/lat space), the covariant path integral can be handled by the usual
methods. In section \ref{s:analytical} we will show that, if the resulting
Lagrangian is a quadratic form, the path integral can be solved
by using well known analytical methods.

\subsection{One-dimensional case} 
 
In the one-dimensional case the covariant formulation
simplifies significantly.
The transformation rule for the metric tensor in 
one dimension is simply
\begin{equation} 
G^\prime_{11}=\left({dx\over dx^\prime}\right)^2 G_{11}, 
\end{equation} 
then $G^\prime_{11}$ can be made equal to $1$ everywhere by 
choosing 
\begin{equation} 
x^\prime=\int dx\;\sqrt{G_{11}}. 
\label{e:ch1} 
\end{equation} 
In this coordinate system the curvature $R$ vanishes,
as follows by Eqs.\ (\ref{e:r2}) and (\ref{e:r3}); as a result,
since $R$ is invariant, it vanishes in every coordinate system. 
Then, by using the notation
$G^{11}=\sigma(x,\tau)^2$, and
$\displaystyle{G_{11}=G={1\over\sigma(x,\tau)^2}}$,
the Lagrangian (\ref{e:cov3}) becomes
\begin{equation} 
{\cal L}[x,{\dot x};\tau]= 
{1\over 2\;\sigma(x,\tau)^2}\; 
[{\dot x}-h(x,\tau)]^2+ 
{\sigma(x,\tau)\over 2}\;{\partial\over\partial x}\; 
{h(x,\tau)\over\sigma(x,\tau)}, 
\label{e:cov3one} 
\end{equation} 
where 
\begin{equation} 
h(x,\tau)=A(x,\tau)-\;{1\over 2}\;\sigma(x,\tau)\;\partial_x\;\sigma(x,\tau). 
\end{equation} 
Finally, the transformation rule of the measure is 
\begin{equation} 
{\cal D}[\sigma^\prime(x^\prime,\tau)^{-1}\;x^\prime(\tau)]=\left| 
{\partial x^\prime\over\partial x}\right|^{-1}_{x,T} 
{\cal D}[\sigma(x,\tau)^{-1}\;x(\tau)].
\label{e:mes2one} 
\end{equation}

\subsection{It\^o lemma} 
 
Let us consider, for the sake of simplicity, the one-dimensional Langevin
equation (\ref{e:1.3}).
The It\^o  lemma states that a function, $z(x)$, of the stochastic variable
follows the process 
\begin{equation} 
\Delta z=\left({\partial z\over\partial x}\;A+{1\over 2}\; 
{\partial^2 z\over\partial x^2}\;\sigma^2\right)\Delta\tau+ 
{\partial z\over\partial x}\;\sigma\;\Delta w. 
\end{equation} 
We can now obtain the It\^o lemma from the path integral 
formalism. 
 
Let us interpret the function, $z(x)$, as a change of 
variables from the variable, 
$x$, to the variable, $z$. The transformation rules for the coefficients of
the Langevin equation can be obtained by 
\begin{eqnarray} 
\sigma^\prime&=&{\partial z\over\partial x}\;\sigma,
\label{e:ito3a}\\
h^\prime&=&{\partial z\over\partial x}\;h, 
\label{e:ito3b} 
\end{eqnarray} 
where 
\begin{eqnarray} 
h(x,\tau)&=&A(x,\tau)-{1\over 2}\;\sigma(x,\tau)\;\partial_x\sigma(x,\tau),\\ 
h^\prime(z,\tau)&=&A^\prime(z,\tau)- 
{1\over 2}\;\sigma^\prime(z,\tau)\;\partial_z\sigma^\prime(z,\tau).
\end{eqnarray}
The Eq. (\ref{e:ito3a}) gives directly the standard deviation
for the new variable, $z$.  Moreover, by using also the relation 
\begin{equation} 
\partial_z\sigma^\prime 
={\partial x\over\partial z}\;{\partial\over\partial x}\; 
({\partial z\over\partial x}\;\sigma)\nonumber={\partial x\over\partial z}\; 
\left({\partial^2 z\over\partial x^2}\;\sigma+{\partial z\over\partial x}\; 
\partial_x\sigma\right), 
\end{equation} 
we obtain 
\begin{equation} 
h^\prime=A^\prime-{1\over 2}\;\sigma\; 
\left({\partial^2 z\over\partial x^2}\;\sigma+{\partial z\over\partial x}\; 
\partial_x\sigma\right). 
\end{equation} 
Finally, if we compare this expression with 
\begin{equation} 
h^\prime={\partial z\over\partial x}\;h={\partial z\over\partial x}\; 
(A-{1\over 2}\;\sigma\;\partial_x\sigma), 
\end{equation} 
we obtain the drift for the variable, $z$, 
\begin{equation} 
A^\prime={\partial z\over\partial x}\;A+ 
{1\over 2}\;{\partial^2 z\over\partial x^2}\;\sigma^2. 
\label{e:ito4} 
\end{equation} 
The expressions (\ref{e:ito3a}) and (\ref{e:ito4}) coincide with the results
of the It\^o lemma.

\section{Expectation values}
\label{s:exp}
 
The expectation value of a functional $g[{\bf x}(\tau);\tau]$
on the stochastic 
process defined by the Langevin equation (\ref{e:1.3corr}), {\em with fixed 
initial and final conditions}, ${\bf x}(t)={\bf y}$,
and ${\bf x}(T)={\bf x}$, is formally given by 
\begin{eqnarray} 
\lefteqn{<{\bf x},T\mid g[{\bf x}(\tau);\tau]\mid {\bf y},t>_L\;=}\nonumber\\ 
&&\;\;\;\;\;\;\;\;\;\;\;\;\;\;{\int\!\!\!\int}_{{\bf x}(t)={\bf 
y}}^{{\bf x}(T)={\bf x}} {\cal D}[\sqrt{G}\;{\bf 
x}(\tau)]\;g[{\bf x}(\tau);\tau]\; 
\exp\left\{-\int_t^T L[{\bf x}(\tau),{\dot 
{\bf x}}(\tau);\tau]\;d\tau\right\}, 
\label{e:1.9aa} 
\end{eqnarray} 
The  discretized expression corresponding to (\ref{e:1.9aa}) is 
\begin{eqnarray} 
&&<{\bf x},T\mid g[{\bf x}(\tau);\tau]\mid {\bf y},t>_L\;\simeq
\int_{-\infty}^{+\infty}\!\!\!\!\!\!\!\!\!\ldots\! 
\int_{-\infty}^{+\infty} 
\prod^{N}_{i=1}dx^n_i \prod^{N+1}_{j=1} 
(2\pi\Delta t)^{-n/2}\;\sqrt{G({\bf x}_{j-1},t_{j-1})}\nonumber\\ 
&&\;\;\;\;\;\;\;\;\;\;\;\;\;\;\;\;\;\;\;\;\;\;\;\;\;\;\;\;\;\;\;\;\;\;\times\; 
g({\bf x}_t,{\bf x}_1,\ldots,{\bf x}_T;t,t_1,\ldots,T)\; 
\exp\left\{-S({\bf x}_j,t_j;{\bf x}_{j-1},t_{j-1})\right\},
\label{e:1.9c} 
\end{eqnarray} 
where $S({\bf x}_j,t_j;{\bf x}_{j-1},t_{j-1})$ represents an appropriate
discretization rule, while the function
$g({\bf x}_t,{\bf x}_1,\ldots,{\bf x}_T;t,\tau_1,\ldots,T)$ is a 
discretization of the functional $g[{\bf x}(\tau);\tau]$. In general, 
different discretizations could give different results.

Obviously, the expectation value (\ref{e:1.9aa}) can be also written in
covariant form by substituting the Lagrangian,
$L[{\bf x}(\tau),{\dot {\bf x}}(\tau);\tau]$, with the invariant one,
${\cal L}[{\bf x}(\tau),{\dot {\bf x}}(\tau);\tau]$, by using the covariant
discretization rule (\ref{e:cov5}), and provided that also the
functional, $g[{\bf x}(\tau);\tau]$, is discretized in a covariant form.

A case of particular importance is when the functional can be written as
\begin{equation} 
g[{\bf x}(\tau);\tau]=\exp\left\{\int_t^T V[{\bf x}(\tau);\tau]\;d\tau\right\}.
\label{e:pot} 
\end{equation}
Since the term, $V[{\bf x}(\tau);\tau]$, is invariant
for a general transformation of variables, it can be simply included into
the Lagrangian both in the non-covariant and the covariant formulations.
Moreover, since such a term does not depend
on the derivative of the stochastic variable, the differences among
different discretizations of the functional (\ref{e:pot}) are of order
larger than $O(\Delta t)$, and, therefore, it can be discretized with any rule.

\subsection{Mean value}

There are several examples of functionals. The simplest one is the expectation 
value of the stochastic variable $x$ at a given time $t_1$. 
The functional can be formally written as
\begin{equation} 
g[x(\tau);\tau] =\int_{-\infty}^{+\infty}\delta(t_1-\tau)x(\tau)d\tau
\;\;\;\;\;\;\;\;\;t\leq t_1\leq T,
\end{equation} 
and the expectation value is given by 
\begin{equation} 
<x,T\mid x(t_1)\mid y,t>_L\;=\int_{-\infty}^{+\infty} 
dz\;\rho(x,T\mid z,t_1)\;z\; 
\rho(z,t_1\mid y,t). 
\end{equation}
For instance, in the simple case given by Eq. (\ref{eq:free})
the expectation value is easy to evaluate, and we obtain
\begin{eqnarray} 
<x,T\mid x(t_1)\mid y,t>_L&=&{1\over{\sqrt{2 \pi\sigma^2\, (T-t_1)}}} 
{1\over{\sqrt{2 \pi\sigma^2\, (t_1-t)}}}\nonumber\\ 
&&\times\;
\int_{-\infty}^{+\infty}\!dz \;
\exp{\left[ -{{(x-z)^2}\over{2\,\sigma^2\, (T-t_1)}}\right]}\; z \;
\exp{\left[ -{{(y-z)^2}\over{2\,\sigma^2\,(t_1-t)}}\right]}.
\end{eqnarray} 
In absence of the integrand factor, $z$, we get the 
transition amplitude from $t$ to $T$,
\begin{equation} 
\rho(x,T\mid y,t)= {1\over{\sqrt{2 \pi\sigma^2\, (T-t)}}} 
\exp{\left[ -{{(x-y)^2}\over{\,2\sigma^2\,(T-t)}}\right]}.
\label{e:previous} 
\end{equation} 
If we perform the integral with 
the integrand factor, $z$, we obtain the transition probability
(\ref{e:previous}) 
multiplied by the mean value of the functional,
\begin{equation} 
<x,T\mid x(t_1)\mid y,t>_L\;={{x (t_1-t) + y (T-t_1)}\over{(T-t)}}\;
\rho(x,T\mid y,t).
\end{equation}

\subsection{Functionals for financial quantities} 

Many financial quantities are defined as expectation value 
of functionals on a stochastic process, {\em with fixed 
initial condition}, ${\bf x}(t)={\bf y}$. We will denote this kind
of expectation value by the symbol, $E_L$, which is defined by
\begin{equation}
E_L\Bigl[g[{\bf x}(\tau);\tau]\mid {\bf x}(t)={\bf y}\Bigr]=
\int^{+\infty}_{-\infty}d^nx
<{\bf x},T\mid g[{\bf x}(\tau);\tau]\mid {\bf y},t>_L.
\label{e:expe}
\end{equation}

Note that this quantity could be formally written as a functional integral
with only one extreme fixed, ${\bf x}(t)={\bf y}$. In this case the
functional measure would be completely invariant, and the whole functional
integral could be formally written in full invariant form. However,
from a computational point of view, this does not essentially simplify the
procedure of calculation.

\subsubsection{Zero-coupon bond}

A first example is the evaluation of the quantity
\begin{equation}
P(r_t,t,T)=E_L\left[e^{-\int_t^Tr(\tau)d\tau}\mid r(t)=r_t\right],
\label{eq:fun}
\end{equation}
which is the  price at time $t$  
of a zero-coupon discount bond with principal 1, and maturing at time $T$.
The variable, $r(\tau)$, is the interest rate, and satisfies a SDE
corresponding to the Lagrangian, $L[r(\tau),{\dot r}(\tau);\tau]$.
Let us define
\begin{eqnarray} 
G(r_T,T\mid r_t,t)&=&<r_T,T\mid e^{-\int_t^Tr(\tau)d\tau}\mid r_t,t>_L\nonumber\\
&=&{\int\!\!\!\int}_{r(t)=r_t}^{r(T)=r_T} 
{\cal D}[\sigma(r,\tau)^{-1}r(\tau)]\;
\exp\left\{-\int_t^T {\tilde  L}[r(\tau),{\dot r}(\tau);\tau]\;d\tau\right\},
\label{eq:gj}
\end{eqnarray} 
where
\begin{equation} 
{\tilde L}[r(\tau),{\dot r}(\tau);\tau]=L[r(\tau),{\dot r}(\tau);\tau]+r(\tau).
\label{eq:lv} 
\end{equation} 
The price is then given by
\begin{equation}
P(r_t,t,T)=\int_{-\infty}^{+\infty}dr_T\;G(r_T,T\mid r_t,t).
\label{e:zcbp}
\end{equation}

\subsubsection{Caplet}

The price of a caplet is defined by the following expectation value:
\begin{equation} 
C(r_t,t,T,s)=E_L\left[e^{-\int_t^Tr(\tau)d\tau}\; (\chi-P(r_T,T,s))\;
\theta\Bigl(\chi-P(r_T,T,s)\Bigr)\mid r(t)=r_t\right],
\end{equation} 
which can be seen as the price at time $t$ of a European put option
maturing at time $T$ on a zero-coupon
discount bond with principal 1, and maturing at time $s$, where $\chi$ is
the strike price.
This expression can be put in the form 
\begin{eqnarray} 
C(r_t,t,T,s)&=&\int_{-\infty}^{+\infty}dr_T{\int\!\!\!\int}_{r(t)=r_t}^{r(T)=r_T} 
{\cal D}[\sigma(r,\tau)^{-1}r(\tau)]\nonumber\\
&&\times\;(\chi-P(r_T,T,s))\;\theta\Bigl(\chi-P(r_T,T,s)\Bigr)
\;\exp\left\{-\!\int_t^T\! \{{\tilde L}[r(\tau),{\dot 
r}(\tau);\tau]\}\;d\tau\right\}\nonumber\\ 
&=&\int_{-\infty}^{+\infty}dr_T\;\;(\chi-P(r_T,T,s))\;
\theta\Bigl(\chi-P(r_T,T,s)\Bigr)\;G(r_T,T\mid r_t,t), 
\label{e:1.27} 
\end{eqnarray} 
where $G(r_T,T\mid r_t,t)$ and ${\tilde L}[r(\tau),{\dot r}(\tau);\tau]$ have been
defined in Eqs. (\ref{eq:gj}) and (\ref{eq:lv}), respectively.
This result has been reported, and explicitly calculated
for the Vasicek model, by Jamshidian (cf. Eq. (11) in Ref.\cite{jam}).

\section{Analytical results}
\label{s:analytical}

In general, the explicit analytical calculation of a path integral is
a formidable task, and it is possible in a very few cases only. A class
of systems which allows exact path integration is characterized by quadratic
Lagrangians\cite{feynman-hibbs,papa,schul}. 
In the next sections, we will consider some cases which belong or can be
reduced to this class.

\subsection{Elementary cases}
 
\subsubsection{Harmonic Lagrangian}

The harmonic Lagrangian is  
\begin{equation} 
L[x,{\dot x};\tau]={1\over 2\sigma^2}\;{\dot x}^2+{\omega^2\over 
2\sigma^2}\;x^2. 
\label{e:2.48} 
\end{equation} 
Note that this Lagrangian does not contain any term with a coupling
between
$x$ and ${\dot x}$, and
cannot correspond to any stochastic process, but it is the starting
point for the next calculations.
The solution of the path integral for this process is a
well known result, and it can be
cast in the form 
\begin{equation} 
I_{harmonic}(x,T\mid y,t)= 
e^{\omega (x^2-y^2)/2\sigma^2}\;e^{-\omega(T-t)/2}\; 
{1\over\sqrt{2\pi{\bar\sigma}^2}}\; 
\exp\left\{-{(y\,e^{-\omega(T-t)}-x)^2\over 
2{\bar\sigma}^2}\right\},
\label{e:2.57} 
\end{equation} 
where 
\begin{equation} 
{\bar\sigma}=\sigma\;\sqrt{(1-e^{-2\omega(T-t)})\over 2\omega}. 
\label{e:2.58} 
\end{equation}

\subsubsection{Harmonic Langevin equation}
\label{s:hle}

Let us consider the stochastic process defined by the following SDE,
\begin{equation} 
\Delta x = -\omega x \Delta t +\sigma \Delta w.
\label{e:2.60} 
\end{equation} 
The corresponding Lagrangian is  
\begin{equation} 
L[x,{\dot x};\tau]={1\over 2\;\sigma^2}\;[{\dot x}+\omega\,x]^2-
{\omega\over 2}. 
\label{e:2.59} 
\end{equation} 
This Lagrangian is equal to the previous one plus a coupling between the
stochastic variable, $x$, and its derivative, ${\dot x}$, and plus a constant
factor. By using the prescription given in section \ref{s:gauge}, we get
the following finite time transition probability,
\begin{equation} 
\rho(x,T\mid y,t)
=e^{-\omega(x^2-y^2)/2\sigma^2}\, 
e^{\omega(T-t)/2}\, 
{\int\!\!\!\int}_{x(t)=y}^{x(T)=x} 
{\cal D}[\sigma^{-1}x(\tau)]\; 
\exp\left\{-\int_t^T 
\left[{{{\dot x}^2}\over 2\sigma^2}+ 
{{\omega^2 x^2}\over 2\sigma^2}\right] 
d\tau\right\}.
\label{e:2.61} 
\end{equation} 
The path integral in Eq. (\ref{e:2.61}) is just $I_{harmonic}(x,T\mid y,t)$.
Then, we obtain 
\begin{equation} 
\rho(x,T\mid y,t)= 
{1\over\sqrt{2\pi{\bar\sigma}^2}}\; 
\exp\left\{-{(y\,e^{-\omega(T-t)}-x)^2\over 
2{\bar\sigma}^2}\right\}, 
\label{e:2.62} 
\end{equation} 
where ${\bar\sigma}$ is given by Eq. (\ref{e:2.58}).

\subsection{The Vasicek model (Ornstein-Uhlenbeck)} 
 
This model is very popular in the financial literature and it is defined by 
the stochastic equation\cite{ornulh}
\begin{equation} 
\Delta r = a(b-r)\Delta t+\sigma\Delta w,
\label{e:2.63} 
\end{equation} 
where the variable, $r$, is the short term interest rate.
The finite time transition probability, which, following the notation
of Ref.\cite{jam}, will be denoted here by
$K(r_T,T\mid r_t,t)$, is given by 
\begin{equation} 
K(r_T,T\mid r_t,t)={\int\!\!\!\int}_{r(t)=r_t}^{r(T)=r_T} 
{\cal D}[\sigma^{-1}r(\tau)]\; 
\exp\left\{-\int_t^T L^{\rm V}[r(\tau),{\dot 
r}(\tau);\tau]\;d\tau\right\},
\label{e:2.65} 
\end{equation}    
where 
\begin{equation} 
L^{\rm V}[r,{\dot r};\tau]={1\over 2\;\sigma^2}\;[{\dot r}-a\,(b-r)]^2-
{a\over 2}. 
\label{e:2.66}
\end{equation} 
 
This case is similar to that of section \ref{s:hle}, with
\hbox{$\omega=a$}, and \hbox{$x\rightarrow r-b$}. Then the
path integral can be simply worked out 
by changing the variables (here we do not need the covariant formulation
since this is a linear transformation of variables).

\subsubsection{Expectation value} 

Actually, we are not really interested in the Green function for
Eq. (\ref{e:2.63}), but in the expectation value (\ref{eq:gj}), i.e.
\begin{equation}
G(r_T,T\mid r_t,t)={\int\!\!\!\int}_{r(t)=r_t}^{r(T)=r_T} 
{\cal D}[\sigma^{-1}r(\tau)]\;
\exp\left\{-\int_t^T {\tilde L}^{\rm V}
[r(\tau),{\dot r}(\tau);\tau]\;d\tau\right\},
\label{eq:gjv}
\end{equation}
where ${\tilde L}^{\rm V}[r,{\dot r};\tau]=L^{\rm V}[r,{\dot r};\tau]+r$.
We can make the following linear change of variable,
\begin{equation} 
z=r-b+{{\sigma^2}\over{a^2}}.
\label{e:2.71}
\end{equation} 
The Jacobian of the transformation is equal to 1, and the Lagrangian becomes
\begin{equation} 
{\tilde L}^{\rm V}[z,{\dot z};\tau]={1\over 2\;\sigma^2}\; 
[{\dot z}+ az)]^2 -{a\over 2} + b -{{\sigma^2}\over{2 a^2}}- 
{{{\dot z}}\over{a}}.
\end{equation}
The last three terms can be integrated, and they give rise to a phase factor
(we recall that we are using the mid-point prescription), then
the integral (\ref{eq:gjv}) can be written as
\begin{equation} 
G(r_T,T\mid r_t,t)=e^{-\Delta\theta} \;
{\int\!\!\!\int}_{z(t)=r_t-b+\sigma^2/a^2}^{z(T)=r_T-b+\sigma^2/a^2}
{\cal D}[\sigma^{-1}z(\tau)]\; 
\exp\left\{-\int_t^T \left[{1\over 2\sigma^2}\,({\dot z}+ 
a\,z)^2-{a\over 2}\right]\;d\tau\right\},
\label{e:2.72}
\end{equation} 
where the phase factor, $\Delta\theta$, is given by  
\begin {equation} 
\Delta\theta=\int_t^T \left( b -{{\sigma^2}\over{2 a^2}}- 
{{{\dot z}}\over{a}}\right)\;d\tau = {{r_T -r_t}\over{a}}+ 
(T-t) \left[ b -{{\sigma^2}\over{2 a^2}}\right].
\end{equation}
Finally, since the remaining path integral is equal to that corresponding to
the harmonic Langevin equation,
by using the result (\ref{e:2.62}) we obtain 
\begin{eqnarray} 
&&G(r_T,T\mid r_t,t)={{e^{{(r_T -r_t)/{a}}+ 
(T-t) \left[ b -{{\sigma^2}/{2 a^2}}\right]}}\over 
\sqrt{2\pi{\bar\sigma}^2}}\nonumber\\
&&\;\;\;\;\;\;\;\;\;\;\;\;\;\;\;\;\;\;\;\;\;\;\;\;\times\;
\exp\left\{-{[(r_t-b+{{\sigma^2}/{a^2}})\,e^{-a(T-t)}- 
(r_T-b+{{\sigma^2}/{a^2}})]^2\over 
2{\bar\sigma}^2}\right\}, 
\label{e:2.73}
\end{eqnarray} 
with
\begin{equation} 
{\bar\sigma}=\sigma\;\sqrt{(1-e^{-2a(T-t)})\over 2a}. 
\label{e:2.74}
\end{equation} 
This expression is the same that we 
find in the literature (cf. Eq. (13) in Ref.\cite{jam}).

\subsubsection {Zero-coupon bond and caplet price in the Vasicek model}

The quantity $G(r_T,T\mid r_t,t)$  allows to evaluate the price,
$P(r_t,t,T)$, at time $t$ of
a zero-coupon discount bond with principal 1, and maturing at time $T$,
by employing the formula (\ref{e:zcbp}).
Since $r_T$ appears in a quadratic form to the exponent, the integral can be 
performed  by completing the square and using the rules for 
Gaussian integration. 
The final result is given by 
\begin{equation} 
P(r,t,T)=\exp \left[ -B(T-t)\; r +[B(T-t)-(T-t)](b-{\sigma^2\over{2a^2}})- 
B(T-t)^2\;{\sigma^2 \over{4a}} \right],
\label{e:2.91} 
\end{equation} 
where
\begin{equation} 
B(\tau) = {{1-e^{-a\tau}}\over a}.
\label{e:2.92} 
\end{equation} 

A second task is to evaluate the caplet price given by Eq. (\ref{e:1.27}).
Such an expression has the structure of a truncated lognormal integration,
and can be done analytically by usual methods.
Here we do not discuss the details of the calculation which are
given in the literature.

\subsection {The Black-Scholes model} 
\label{black} 
 
Let us consider the Black-Scholes model 
\begin{equation} 
\Delta S=a\;S\;\Delta \tau+\sigma\;S\;\Delta w. 
\label{e:bs1} 
\end{equation} 
where $S$ is the stock price.
Here the metric depends on the stochastic variable, $S$, and we need a
non-linear change of variables to perform the calculation. Therefore we must
use the covariant formulation.

The invariant Lagrangian corresponding
to Eq. (\ref{e:bs1}) can be obtained by
Eq. (\ref{e:cov3one}), with $\displaystyle{\sigma(S)=\sigma\; S}$, and 
$\displaystyle{h(S)=a\;S-{\sigma^2\over 2}\;S}$. Since this is a 
\hbox{one-dimensional} problem, we can use the change of variables of
Eq.\ (\ref{e:ch1}) which gives rise to a constant metric. Then the
relation between the two coordinate systems is
\begin{equation} 
S^\prime=\int {dS\over S}=\log S. 
\label{e:ch2} 
\end{equation} 
Since the Lagrangian is invariant, we only need to exchange the quantities in
the old coordinate system with the new ones, given by
\begin{eqnarray} 
\sigma^\prime(S^\prime)&=&\sigma,\\ 
h^\prime(S^\prime)&=&{dS^\prime\over dS}\;h(S)= 
(a-{\sigma^2\over 2}).
\end{eqnarray} 
The invariant Lagrangian becomes 
\begin{equation} 
{\cal L}[S^\prime,{\dot S}^\prime;\tau]={1\over 2\sigma^2}\; 
[{\dot S}^\prime-(a-{\sigma^2\over 2})]^2.
\label{e:black}
\end{equation} 
Furthermore, the new measure is
\begin{equation} 
{\cal D}[\sigma^{-1}S^\prime(\tau)]=S_T\;
{\cal D}[\sigma(S)^{-1}S(\tau)]. 
\label{e:mblack} 
\end{equation} 
Since in the new coordinate system the invariant Lagrangian (\ref{e:black})
coincides with the mid-point one
(see, section \ref{s:gconst}), we can proceed in the usual way, and we get for
the finite time transition probability,
\begin{eqnarray} 
&&\rho(S_T,T\mid S_t,t)=\nonumber\\
&&\;\;e^{\Delta\phi/\sigma^2}\;{1\over S_T}\; 
{\int\!\!\!\int}_{S^\prime(t)=\log(S_t)}^{S^\prime(T)=\log(S_T)}
{\cal D}[\sigma^{-1}S^\prime(\tau)]\; 
\exp\left\{-\int_t^T 
\left[{1\over 2\sigma^2}\,{\dot S}^{\prime\,2}+ 
{1\over 2\sigma^2} \left({a-{\sigma^2\over 2}}\right)^2 \right] 
d\tau\right\}, 
\label{e:2.106} 
\end{eqnarray} 
where 
\begin{equation} 
\Delta\phi=\int_{\log(S_t)}^{\log(S_T)} ({a-{\sigma^2\over 2}})\; dS^\prime = 
({a-{\sigma^2\over 2}})\; \Bigl[\log(S_t)- \log(S_T)\Bigr].
\label{e:2.107} 
\end{equation} 
Finally, by using the formulae obtained previously, and collecting
these expressions, we get the result of Black and Scholes.

\subsection{The Cox Ingersoll Ross model}
 
Another case where we need the covariant formulation of path integrals is
the Cox Ingersoll Ross (CIR) model.
This model has been developed to overcome a limit
of the Vasicek model, i.e. the fact that the short term interest rate can
take negative values. Obviously, this fact has not any real meaning.
To avoid this problem Cox, Ingersoll, and Ross\cite{cir} have proposed
the following process,
\begin{equation}
\Delta r = a\,(b-r)\;\Delta\tau+\sigma\; \sqrt r\;\Delta w
\label{e:2.108}
\end{equation}
(for a discussion of this equation, see, for example, Ref.\cite{hull}).
Here the variance is proportional to $r$, and goes to zero with the short
term interest rate.
The choice of this particular dependence was related to the
solubility of the model.
As we will see, in this case the path integral (\ref{eq:gj}) can be
expressed in a closed form through an analytical function.

Let us start from the path integral written in covariant
form. The invariant Lagrangian is given by the expression
(\ref{e:cov3one}), with
$\displaystyle{\sigma(r)=\sigma\;\sqrt{r}}$,
and $\displaystyle{h(r)=a\;(b-r)-{\sigma^2\over 4}}$.
If we introduce the variable $z = 2\sqrt r $, we obtain
\begin{eqnarray}
\sigma^\prime(z)&=&\sigma,\\
h^\prime(z)&=&{dz\over dr}\;h(r)=
2\;\left({a(b-z^2/4)\over z}-{\sigma^2\over 4\,z}\right).
\end{eqnarray}
Therefore, in the new variable, the invariant Lagrangian becomes
\begin{equation}
{\cal L}[z,{\dot z};\tau] ={1\over {2 \sigma^2}}\left[{\dot z} -
2\left({a(b-z^2/4)\over
z}-{\sigma^2\over {4\,z}}\right) \right]^2+
{1\over 2}\;{\partial\over\partial z}\,h^\prime(z).
\label{e:2.113}
\end{equation}
In order to evaluate the
expectation value (\ref{eq:gj}), we must add to the Lagrangian above the term,
$r= z^2/4$. Furthermore, the measure in the new variable is given by
\begin{equation}
{\cal D}[\sigma^{-1}z(\tau)]=\sqrt{r_T}\;
{\cal D}[\sigma(r)^{-1}r(\tau)].
\label{e:mcir}
\end{equation}
If we now observe that
the term proportional to ${\dot z}$ gives rise to the following phase factor
(see, section \ref{s:gauge}),
\begin{equation}
\Delta\phi = \int_{z_t}^{z_T} h^\prime(z)\;dz=
-{a\over{4}}(z_T^2-z_t^2) +
(2ab-{\sigma^2\over 2}) \log{{z_T}\over{z_t}},
\end{equation}
with $z_T=2\sqrt{r_T}$, and $z_t=2\sqrt{r_t}$,
then the expectation value (\ref{eq:gj}) can be written as
\begin {equation}
G(z_T,T\mid z_t,t)={2\over z_T}\;
\exp\left\{{\Delta\phi\over\sigma^2}\right\}\;
{\int\!\!\!\int}_{z(t)=z_t}^{z(T)=z_T} 
{\cal D}[\sigma^{-1}z(\tau)]\;
\exp\left\{-\int_t^T L_{\rm eff}[z,{\dot z};\tau]\;d\tau\right\},
\label{e:pathcir}
\end {equation}
where
\begin {equation}
L_{\rm eff}[z,{\dot z};\tau]={1\over {2 \sigma^2}}\;{\dot z}^2+
c_1 + c_2\, z^2 + { c_3\over z^2},
\label{e:2.115}
\end {equation}
and the parameters, $c_1$, $c_2$, and $c_3$, are given by
\begin {eqnarray}
c_1 &=& -{a^2b\over{\sigma^2}}, \\
c_2 &=&{a^2\over{8\sigma^2}}+{1\over 4},\\
c_3 &=&{(4ab-3\sigma^2)\;(4ab-\sigma^2)\over 8\sigma^2}.
\label{e:2.116}
\end {eqnarray}
The effective Lagrangian, $L_{\rm eff}$, is not quadratic, but the term
$\displaystyle{\sim{1\over z^2}}$ is the same as the centrifugal contribution
in the case of a two-dimensional harmonic Lagrangian, written in polar
coordinates. Such a case can be solved in arbitrary dimension.
The explicit expression and the algebraic
and analytical manipulations of these rather cumbersome mathematical aspects
are given in Appendix \ref{a:c}.

The solution of the path integral (\ref{e:pathcir}) can be used to
calculate the zero-coupon price by Eq. (\ref{e:zcbp}).
The final result, quoted in the literature (see, for example,\cite{hull}), can
be written as
\begin{equation}
P(r_t,t,T) = A(t,T)\; e^{-\,B(t,T)\,r_t},
\label{e:phull}
\end{equation}
where the functions $B(t,T)$, and $A(t,T)$ are given in the Eqs. (\ref{e:btt}),
and (\ref{e:att}).
We stress further that in this case the anomalous $1/z^2$ dependence masks the
underlying quadratic problem. Once this connection is understood the solution
is straightforward.

The expression above represents a closed-form solution for the zero-coupon
price in the CIR model.
Obviously, the explicit knowledge of the Green function allows the evaluation
of more complex functionals such as, for example, cap or floor prices.
 
Moreover, the method can be applied to coupled SDEs.
The general theorems on quadratic Lagrangians allow to solve by
diagonalization a multi-dimensional coupled problem, the only constraint is
the absence of higher order terms ($O(x^3)$ or larger).

Finally, this treatment can be extended to more complex problems
with non-quadratic Lagrangians.
The scientific literature suggests a variety of
approximation techniques: for example, the perturbative, and the saddle point
methods.

In conclusion, the path integral approach, besides other analytical
methods (see, for example, \cite{heston}), represents a powerful tool of
analysis.

\section{Conclusions}

We have described the path integral method as an alternative approach to find
the solution of stochastic equations.
We have shown that the method can be used in general cases, and that it is
equivalent to the formulation in terms of SDE and PDE equations.
The method is suitable in particular for defining functional mean values and
path dependent problems.
However, an exact analytical treatment is possible only in a few cases. A
constraint necessary to get analytical solutions is the quadratic form of
the Lagrangian describing the stochastic process.
We have discussed in general this case, and the problem of the
coupling between the stochastic variable and its derivative.
Moreover, we have given the explicit solution for some important
one-dimensional problems, and briefly discussed the
generalization to quadratic multi-dimensional cases.
Finally, we point out that
the path integral treatment can be extended to non-quadratic
cases by approximate analytical techniques, or by numerical methods.
The numerical approach will be described in a forthcoming paper
(see also Ref.\cite{rctlecture}).

\appendix 

\section{Equivalence between the path integral approach and the partial
derivative equations}
\label{a:a}

We want to show that the short-time transition probability (\ref{e:onepre})
gives rise to the Fokker-Planck equation (\ref{e:1.5a})
corresponding to the differential stochastic equation (\ref{e:1.3}).
Let us write the following identity
\begin{equation} 
\rho(x,T+\Delta t\mid y,t)=
\int_{-\infty}^{+\infty}dz\;\rho(x,T+\Delta t \mid z,T)\;
\rho (z,T\mid y,t).
\label{e:prod}
\end{equation}
If we use the approximate expression (\ref{e:onepre}), and employing the
identity $z=x+(z-x)=x+\eta$, we obtain
\begin {equation} 
\rho(x,T+\Delta t\mid y,t)\simeq \int_{-\infty}^{+\infty}d\eta\;
{1\over{\sqrt{ 2\pi\Delta t\, \sigma^2}}} \;
\exp\left\{ {{(-\,\eta- A(x+\eta,T)\,\Delta t)^2}\over{2\,\Delta 
t\,\sigma^2}}\right\}\rho(x+\eta,T\mid y,t),
\end{equation} 
where, for the sake of simplicity, we have taken
$\sigma={\rm constant}$.
If we recall that $\eta\sim O(\sqrt{\Delta t})$, and expand the above
expression in $\eta$ up to $O(\Delta t)$, we get
\begin{eqnarray} 
\lefteqn{\rho(x,T+\Delta t\mid y,t)\simeq\int_{-\infty}^{+\infty}{{d\eta}
\over{\sqrt{ 2\pi\Delta t \,\sigma^2}}}\;
{\exp\left\{-{\eta^2\over {2\,\Delta t\, \sigma^2}}\right\}}}\nonumber\\
&&\;\;\;\;\;\;\;\;\;\;\;\;\times\;\left\{1- {{\eta}\over {\sigma^2}}\; A(x,T)
- {\eta^2\over\sigma^2}\;{\partial A(x,T)\over\partial x}+ 
{\eta^2\over{\sigma^4}}\;A(x,T)^2\right\}
\left\{1-{{\Delta t\, A(x,t)^2}\over{2\sigma^2}}\right\}\nonumber\\
&&\;\;\;\;\;\;\;\;\;\;\;\;\times\;\left\{ 
\rho(x,T\mid y,t) + \eta\;{\partial\over {\partial x}}\,\rho(x,T\mid y,t)
+{\eta^2\over 2}\;
{\partial^2\over {\partial x^2}}\,\rho(x,T\mid y,t)\right\}.
\end{eqnarray} 
Then, by performing all calculations, we find
\begin {equation} 
\rho(x,T+\Delta t\mid y,t)\simeq \rho(x,T\mid y,t) -\Delta t\,\left[
{{\partial\, A(x,T)\, \rho(x,T\mid y,t)}\over {\partial x}} +  
{1\over 2}\; {{\partial^2\, \sigma^2\, \rho(x,T\mid y,t)}
\over {\partial x^2}}\right].
\end{equation}
Since the L.H.S. of this equation can be also written, up to $O(\Delta t)$, as
\begin {equation} 
\rho(x,T+\Delta t\mid y,t)\simeq \rho(x,T\mid y,t) +
\Delta t\; {{\partial }\over\partial T}\,\rho(x,T\mid y,t),
\label{e:expdt}
\end{equation}
by collecting all pieces together, we obtain the Fokker-Planck equation.

The
general case where there is an explicit dependence of the function, $\sigma$,
on $x$, and $T$, is more complex. We must expand
the equation up to $O(\eta^8)$, and $O(\Delta t^4)$. However after several 
cumbersome calculation we obtain the final result  
\begin {equation} 
{{\partial\rho(x,T\mid y,t)}\over{\partial T}} = -{{\partial\, A(x,t)\,
\rho(x,T\mid y,t)}\over {\partial x}} +
{1\over 2} {{\partial^2\, \sigma^2(x,t)\,
\rho(x,T\mid y,t)}\over {\partial x^2}},
\end{equation} 
which coincides with the general Fokker-Planck equation (\ref{e:1.5a}).

\section{Exponent expansion}
\label{a:b}

In this appendix we describe a procedure to obtain an expression for the
short-time transition probability which is correct up to any given
order\cite{makri}.
Here, for the sake of simplicity, we consider the one-dimensional case,
but the formalism can be easily extended to arbitrary dimensions.
Moreover we take $\sigma = {\rm constant}$, and $A(x,\tau)=a(x)$.

The starting point is that the transition probability must satisfy the
Fokker-Planck equation (\ref{e:1.5a}). Therefore, if we make the following
ansatz,
\begin{equation} 
\rho(x ,t+\Delta t\mid y,t)={1\over\sqrt{2\,\pi\,\Delta t\,\sigma^2}}\; 
\exp\left\{{-(x-y)^2\over 2\,\Delta t\,\sigma^2} -f(x,y,\Delta t)\right\},
\end{equation} 
the function, $f(x,y,\Delta t)$, must satisfy the equation 
\begin{eqnarray} 
{\partial\over\partial \Delta t}\;f(x,y,\Delta t)&=&
-a(x)\,{\partial\over\partial x}\;f(x,y,\Delta t)+ 
{1\over 2}\,\sigma^2\,{\partial^2\over\partial x^2}\;f(x,y,\Delta t)
\nonumber\\ 
&&-{\partial\over\partial x}\;a(x) + 
{(x-y)\over \Delta t}{\partial\over\partial x}\; 
f(x,y,\Delta t)- 
{(x-y)\over {\Delta t\,\sigma^2}}\;a(x).
\label{e:ffokk}
\end{eqnarray} 
Let us now expand  $f(x,y,\Delta t)$ in powers of $\Delta t$,
\begin{equation} 
f(x,y,\Delta t)=f_0(x,y)+\Delta t\, f_1(x,y)+\Delta t^2\, f_2(x,y)+ 
\Delta t^3\, f_3(x,y)+\ldots;
\end{equation} 
then, by substituting in (\ref{e:ffokk}), we get the following set of recursive
equations,
\begin{equation} 
\left\{ 
\begin{array}{lcccccc} 
{\partial_x f_0(x,y)}& = &-\displaystyle{1\over\sigma^2}\; a(x)&&&&\\
f_1(x,y)& = & -(x-y)\;{\partial_x}f_1(x,y) & +& 
{1\over 2}\;{\partial_x a(x)} &+&
\displaystyle{a(x)^2\over {2 \sigma^2}}\\
2\, f_2(x,y) &= & -(x-y)\;{\partial_x} f_2(x,y)\; &+ &
{1\over 2}\,\sigma^2\,{\partial^2_x}f_1(x,y)&&\\
3\, f_3(x,y) &= & -(x-y)\;{\partial_x} f_3(x,y) &+ &
{1\over 2}\,\sigma^2\,{\partial^2_x}f_2(x,y)&- &
{1\over 2}\,\sigma^2\,({\partial_x}f_1(x,y))^2\\
4\, f_4(x,y) &= & -(x-y)\;{\partial_x} f_4(x,y) &+ &
{1\over 2}\,\sigma^2\,{\partial^2_x}f_3(x,y)&- &
{1\over 2}\,\sigma^2\,{\partial_x}f_1(x,y)\;
{\partial_x}f_2(x,y)\\
5\, f_5(x,y) &= &\ldots&&&&
\end{array} \right. 
\end{equation} 
The first equation is decoupled by the others and defines an overall 
phase factor. The term $f_0(x,y)$ is related to the integral
of $a(x)$, and it is just the phase factor (\ref{e:1.ex11}).
The second equation gives the first order approximation in $\Delta t$;
its value can be used as input for the successive equation, and so on.
Therefore, by a simple iteration, we can compute the function 
$f(x,y,\Delta t)$ up to the required order. In general,
the structure of these equations is
\begin{equation} 
n f_n(x,y)  =   -(x-y)\;{\partial_x}\, f_n(x,y) +W_n(x,y),
\end{equation} 
where $ W_n(x,y)$ depends on the functions,
$f_1(x,y),\ldots,f_{n-1}(x,y)$. The solution is given by
\begin{equation} 
f_n(x,y)  = \int_{0}^{1} d\xi\; \xi^{n-1}\; W_n(y+ \xi (x-y),y).
\label{e:aa} 
\end{equation} 
Actually the integral appearing in Eq. (\ref{e:aa}) can be rather complicated,
but the problem could be simplified by a further expansion in
$\Delta x=x-y$\cite{rostadwig}.

\section{Explicit calculation for the CIR model}
\label{a:c}

The path integral (\ref{e:pathcir}) has an explicit solution given by
(see, for example, \cite{schul})
\begin {eqnarray}
&&G(z_T,T\mid z_t,t)={2\over z_T}\;
\exp\left\{{\Delta\phi\over\sigma^2}\right\}\;
{\gamma\;\sqrt{z_T\,z_t}\over
2{\sigma^2 \sinh\left[\displaystyle{\gamma\over 2} (T-t)\right]}}\nonumber\\
&&\;\;\;\;\;\;\;\times\;
\exp\left[-{\gamma\over{4\sigma^2}}(z_T^2+z_t^2)
\coth\left(\displaystyle{\gamma\over 2} (T-t)\right)\right]\;
I_{\mu} \left({z_T z_t\;\gamma\over{2\sigma^2
\sinh\left(\displaystyle{\gamma\over 2} (T-t)\right)}}
\right)\; e^{{{a^2b}(T-t)/{\sigma^2}}},
\label{e:2.117}
\end {eqnarray}
with $I_{\mu}(x)$ the modified Bessel function of index $\mu$,
\begin {equation}
\gamma= \sqrt{a^2+2\sigma^2},
\end{equation}
and
\begin{equation}
\mu = {1\over 2}\; \sqrt{1+{{8\, c_3}\over{\sigma^2}}}  =
{{2ab}\over{\sigma^2}}-1.
\end {equation}

In order to get the formula for the zero-coupon price this expression must be
integrated over the final variable, $r_T$:
\begin{equation}
P(r_t,t,T) =
{\int}_{0}^{+\infty}dr_T\;G(z_T,T\mid z_t,t)=
{\int}_{0}^{+\infty}dz_T\;{z_T\over 2}\;G(z_T,T\mid z_t,t).
\end{equation}
The integral looks rather formidable, however it can be handled thanks to the
integral
\begin {equation}
{\int}_{0}^{+\infty} e^{-\alpha z^2} z^{\mu+1}\; I_{\mu}(\beta z)\; dz=
{{\beta^{\mu}}\over{(2\alpha)^{\mu+1}}}\; \exp\left[{\beta^2\over 4\alpha}
\right].
\label{e:2.119}
\end {equation}
In the equation above the parameter $\alpha$, and $\beta$ are given by
\begin{equation}
\alpha={{\gamma}\over{4 \sigma^2}}\, \coth\left[{\gamma\over 2}(T-t)\right]
+{a\over 4 \sigma^2},
\end{equation}
and 
\begin{equation}
\beta= {\gamma z_t\over 2 \sigma^2\,
\sinh\left[\displaystyle{\gamma\over 2}(T-t)\right]}.
\end{equation}
Therefore, by performing the integration, and inserting the original
variable, $r_t$, we obtain the expression
\begin{eqnarray}
P(r_t,t,T) &=& \exp\left[{{\beta^2}\over{4 \alpha}}\right] \;
\exp\left\{{{a\, r_t}\over{\sigma^2}}-{{\gamma}\,r_t\over{\sigma^2}}\coth
\left[\displaystyle{\gamma
\over 2}(T-t)\right]\right\}\nonumber\\
&&\;\;\;\;\times\;\;\left\{
{\gamma\over{4\,\sigma^2\,\alpha\;\sinh\left[\displaystyle{\gamma
\over 2}(T-t)\right]}}\right\}^{\mu+1}\;
{\exp\left[\displaystyle{a^2 b\over\sigma^2} (T-t)\right]}
\end{eqnarray}
Note that
the dependence of $P(r_t,t,T)$ on $r_t$ is only due to the exponential,
$\exp [-B(t,T)\;r_t]$, where
\begin{equation}
B(t,T)= {-\;\gamma^2\over{4\,\alpha\,\sigma^4\;\sinh^2
\left[\displaystyle{\gamma
\over 2}(T-t)\right]}} - {{a}\over{\sigma^2}}+
{{\gamma}\over{\sigma^2}}\coth\left[{\gamma\over 2}(T-t)\right]=
{2({e^{\gamma (T-t)} -1})\over{(\gamma +a)
(e^{\gamma (T-t)} -1) +2 \gamma}}.
\label{e:btt}
\end{equation}
Finally, if we define $A(t,T)$ by
\begin{eqnarray}
A(t,T)&=& \left\{
{\gamma\over{4\,\sigma^2\,\alpha\;\sinh\left[\displaystyle{\gamma
\over 2}(T-t)\right]}}\right\}^{\mu+1}\;
{\exp\left[\displaystyle{a^2 b\over\sigma^2} (T-t)\right]}\nonumber\\
&=&\left [{{2\;\gamma\; e^{\gamma (T-t) /2}}\over{(\gamma +a)
(e^{\gamma (T-t)} -1) +2 \gamma}}\right]^{2ab/\sigma^2}
e^{a^2b (T-t)/\sigma^2},
\label{e:att}
\end{eqnarray}
we can cast the expression for the zero-coupon price, $P(r_t,t,T)$, in the
form given in Eq. (\ref{e:phull}).

\end{document}